\documentclass[pra,twocolumn,showpacs,preprintnumbers,superscriptaddress]{revtex4}
\usepackage{times}
\usepackage{bm}
\usepackage{graphicx}
\usepackage{amsbsy}
\usepackage{amsmath}
\usepackage{amsfonts}
\usepackage{amsthm}
\usepackage{xcolor}
\usepackage{tikz}
\usepackage[utf8]{inputenc}

\begin{document}
	
	\theoremstyle{plain}
	\newtheorem{theorem}{Theorem}
	\newtheorem{lemma}[theorem]{Lemma}
	\newtheorem{corollary}[theorem]{Corollary}
	\newtheorem{proposition}[theorem]{Proposition}\newtheorem{conjecture}[theorem]{Conjecture}
	\theoremstyle{definition}
	\newtheorem{definition}[theorem]{Definition}
	\theoremstyle{remark}
	\newtheorem*{remark}{Remark}
	\newtheorem{example}{Example}
	\title{Classification of three-qubit genuine entangled states using concurrence fill}
	\author{Shruti Aggarwal}
	\email{shruti_phd2k19@dtu.ac.in, 17shruti.a@gmail.com} \affiliation{Delhi Technological University, Shahbad Daulatpur, Main Bawana Road, Delhi-110042,
		India}

	\begin{abstract}
Despite the successful experimental generation and verification of genuine multipartite entanglement, several existing entanglement measures remain insufficient to reliably capture its presence. In this study, we overcome this challenge by utilizing a geometric measure known as \textit{concurrence fill}, which quantifies genuine multipartite entanglement of pure states using the area associated with the underlying concurrence triangle. Firstly, the concurrence fill  for three-qubit pure states is reformulated using three tangle and partial tangles.  This yields a representation that is more tractable and operational. Using this formulation, we derive a criterion to classify GHZ and W class of states.
Next, we analyse the rank-2 mixture of generalized GHZ and W class of states for which three-tangle is known to vanish over a zero polytope. Furthermore, we derive the concurrence fill for the eigenstates of the corresponding mixture and obtain its upper bound for mixed states with zero tangle.
\end{abstract}
\maketitle

\section{Introduction}
Quantum entanglement is an important physical resource in quantum information theory, which distinguishes between quantum and classical worlds. Entanglement has applications in bipartite as well as multipartite quantum systems. Genuine multipartite entanglement (GME) is a significant concept for multipartite systems, since it plays an essential role in quantum communication \cite{karlson1998, yeo2006} and quantum cryptography \cite{hillery1999}.
 Recognizing the significance of entanglement as a valuable resource, it becomes natural to explore its quantification through theoretical methods. For the bipartite system, various entanglement measures have been developed, such as concurrence \cite{wootters1998,wootters1997} , negativity \cite{vidal2002} , entanglement of formation \cite{wootters1998} , etc. Carvalho et al. derived the generalization of concurrence, namely the generalized multipartite concurrence \cite{carvalho2004}. 
 A novel entanglement monotone for GME is proposed, serving as a natural extension of the negativity and coinciding with it in the bipartite case \cite{jungnitch2011}. Additionally, other bipartite entanglement measures, including the relative entropy of entanglement and the robustness of entanglement, are shown to be readily adaptable to the multipartite framework \cite{vidal2001}.\\
Acin et al introduced the classification of mixed three-qubit states into different entanglement classes \cite{acin2001,acinprl2001}. These states are classified into separable, biseparable, and genuine entangled states. There are two different kinds of genuine entangled states, namely, GHZ and W class of states \cite{dur2000}.
Any nontrivial tripartite entangled state can be converted,
by means of stochastic local operations and classical communication (SLOCC), into one of two standard forms, namely, GHZ or W state, which are given as follows:
\begin{eqnarray}
	|GHZ\rangle &=& \frac{1}{\sqrt2} \left(|000\rangle + |111\rangle \right),\\
	|W\rangle &=& \frac{1}{\sqrt3} \left( |100\rangle + |010\rangle + |001\rangle \right).
\end{eqnarray}
GHZ and W class of states are stochastically inequivalent classes of entangled states, i.e., a state in GHZ class cannot be converted into W  class by SLOCC \cite{dur2000}.\\
  A tripartite measure of entanglement known as three-tangle (or tangle) was introduced by Coffman, Kundu and Wootters (CKW) as residual entanglement \cite{ckw2000}. Tangle was derived as consequence of 
 the most significant property of the quantum entanglement, i.e., its monogamous behavior. It plays a significant role in the quantum cryptography which restricts the information gain by an eavesdropper \cite{qaisar2017}, and in quantum channel discrimination \cite{sen2016}. Monogamy relations were first established by considering a three-qubit
 system \cite{ckw2000}. It was shown that the entanglement between the subsystems $A$ and $B$ and entanglement between the subsystems $A$ and $C$ cannot be greater than the entanglement between the subsystems $A$ and $BC$. The concept of monogamy property has been extended to the general $N$-qubit
 system \cite{verstrate2006}. Monogamy inequalities for various entanglement measures such as concurrence \cite{ckw2000}, entanglement of formation \cite{olive2014}, negativity \cite{fan2007,kim2009}, Tsallis-q entanglement
 \cite{kim2010, luo2016} and  Renyi-$\alpha$ entanglement \cite{sanders2010} have
 been further studied. \\
 Quantification of entanglement for multipartite systems is a great challenge due to the much richer mathematical structures involved as compared to the bipartite case. A valid measure of multipartite entanglement must meet two key criteria to be considered a GME measure. These criteria, identified by Ma et al. \cite{ma2011}, are: (a) the measure must be zero for all product and biseparable states, and (b) the measure must be positive for all nonbiseparable states (GHZ and W classes for the three-qubit case).  In \cite{eberly2021}, authors used the monogamy inequality on entanglement distribution in three-qubit systems to introduce a new measure of genuine tripartite entanglement called concurrence fill. It is a GME measure since it is non-zero for GHZ as well as W class of states. The concurrence fill was studied in the experiments of three-flavor neutrino oscillations \cite{li2022}. The extension of concurrence fill to four partite systems, is presented in \cite{mishra}, which represents concurrence fill as a combination of areas of cyclic quadrilateral and triangle structures, arising from two distinct types of bipartitions.\\
Although concurrence fill reflects rich geometric aspects of entanglement and bipartite concurrences, and serves as a useful GME measure, it has its own limitations. Ge et al have shown with the help of an example that concurrence fill is not monotonic under local operations and classical communication (LOCC) \cite{Ge2023}. Interestingly, we have reformulated the concurrence fill in terms of tangle and partial tangles. Since, it has also been proved that the three-tangle is LOCC mononotic, it adds to the relavance of our work \cite{Ge2023}. Moreover, Xie and Eberly derived the concurrence fill for pure three-qubit states \cite{eberly2021}. However, a closed formula for the mixed states GME including that for the concurrence fill in the case of mixed states remains unknown.
 In \cite{njp2008}, the authors present an explicit expression for the three-tangle corresponding to a generalized mixture of GHZ and W classes of states. It has been shown that, for a certain range of the mixing parameter, the three-tangle of a mixed state vanishes. Consequently, a significant subset of states within this family cannot be characterized using the three-tangle, as it fails to detect their entanglement. This limitation adds to the motivation of our study: we aim to quantify the entanglement present in the mixed states $\rho(p)$ with vanishing three-tangle. To this end, we employ the concurrence fill as an alternative measure to estimate the entanglement in these states.\\
   In our work, we study the concurrence fill for pure and mixed three-qubit states.  Firstly, we provide an analytic reformulation of concurrence fill in terms of three-tangle and partial tangle. This formulation is constructive because it has been shown in literature that tangle as well as partial tangles can be measured experimentally \cite{datta2018}. Since tangle is zero for W class of states, it cannot classify W class and GHZ class of states. To overcome this problem we use our formulation of concurrence fill to devise a criterion that classifies GHZ and W class of states in three-qubit systems. In \cite{eberly2021}, authors have introduced the concurrence fill for three-qubit pure states. A corresponding measure for mixed states can be obtained using convex roof extension. This method involves minimizing the concurrence fill of a mixed state $\rho$ over all possible pure state decompositions of $\rho$. Hence, convex roof extension is a powerful but computationally challenging tool. In this work, we consider the family of rank-2 mixture of generalized GHZ and W states in three qubit system. Since all the elements in a decomposition  are superpositions of the eigenstates of the
   given density matrix, the minimum value of concurrence fill will depend on all the  possible concurrence fill values for those superpositions. We derive the concurrence fill for the eigenstates of the density matrices belonging to this family of rank-2 mixed states. Next, we deduce a lower bound of the concurrence fill of these eigenstates. Moreover, we derive an upper bound for the concurrence fill of mixed states with vanishing tangle in this family of states. \\
In section II, few entanglement measures such as concurrence, tangle and concurrence fill have been discussed. Section III contains the reformulation of concurrence fill in terms of tangle and partial tangle followed by criteria for classification of GHZ and W class of states. In section IV, we study the concurrence fill for rank-2 mixed states. Finally, we conclude in section V.

\section{Entanglement Measures}

Several entanglement measures have been proposed, particularly for the bipartite states. Yet, the situation is much more complicated for multipartite systems. If an $N$-partite state is neither separable nor biseparable, it is genuine $N$-partite entangled.
 Any genuine multipartite entanglement measure, denoted by $E(\rho)$ for the state $\rho$ satisfies the following necessary conditions \cite{vedral}.
\begin{itemize}
	\item[{(i)}] $E(\rho) = 0$ if and only if $\rho$ is separable or biseparable.
	\item[{(ii)}] $E(\rho) > 0$ for all genuinely entangled states. 
	\item[{(iii)}] $E(\rho)$ is independent under local unitary transformations, i.e., 
	\begin{equation}
		E(\rho) = U \rho U^{\dagger}
	\end{equation}
where $U$ is an arbitrary local unitary operator.
 \item[{(iv)}]  $E(\rho) \geq \sum_i p_i E(\rho_i)$ where $\{p_i, \rho_i\}$ represents the ensemble of states obtained by applying LOCC operations on $\rho$.
\end{itemize} 

Firstly, let us investigate some entanglement measures that are closely related to concurrence \cite{wootters1998,wootters1997,ckw2000}.
\subsection{Concurrence}
Concurrence is a well-defined measure of entanglement \cite{wootters1998}. Let $\rho_{AB}$ be a density matrix describing a two-qubit state. Then the spin flipped density matrix $\tilde{\rho}_{AB}$ is defined as
\begin{eqnarray}
	\tilde{\rho}_{AB} = (\sigma_{y} \otimes \sigma_{y}) \rho_{AB}^* (\sigma_{y} \otimes \sigma_{y}) \label{rholtilde}
\end{eqnarray}
where $*$ denotes complex conjugation in the standard basis $\{|00\rangle, |01\rangle, |10\rangle, |11\rangle\}$; and $\sigma_y$ is the Pauli matrix given as
\begin{eqnarray}
	  \sigma_y=
	\begin{pmatrix}
		0 & -i \\ 
		i & 0
	\end{pmatrix}
\end{eqnarray}
The concurrence of the state $\rho_{AB}$ is defined as 
\begin{eqnarray}
	C_{AB} =max\{\lambda_1 - \lambda_2 - \lambda_3 - \lambda_4, 0\} \label{concmixed}
\end{eqnarray}
where $\lambda_i's$ are the real eigenvalues of the positive semi-definite matrix ${\rho}_{AB}\tilde{\rho}_{AB}$.\\
If $\rho_{AB}$ is a pure state, its concurrence is reduced to a simple expression as follows \cite{ckw2000}:
\begin{eqnarray}
	C_{AB} = 2 \sqrt{det (\rho_A)}
\end{eqnarray}
where $det(.)$ represents the matrix determinant; and $\rho_A$ is the reduced state (trace of $\rho_{AB}$ over the qubit $B$).\\
There are several alternative formulations of concurrence for bipartite quantum states \cite{hororev2009,aggarwal2021}. These approaches may lead to distinct generalizations for multipartite systems \cite{ligao2024,zhu2020}. These generalizations offer insight into different facets of entanglement, for instance, how different subsystems are entangled with one another, or how entanglement is distributed within a single subsystem \cite{cornelio2013}.

\subsection{Tangle}
Tangle, or three-tangle, extends the idea of concurrence to tripartite systems. For a tripartite pure state $|\psi_{ABC}\rangle$, tangle is expressed as 
\begin{eqnarray}
\tau(|\psi_{ABC}\rangle ) = C^2_{A|BC} - C^2_{AB} - C^2_{AC} \label{tangledef}
\end{eqnarray}
where $C^2_{A|BC}$ represents the amount of bipartite entanglement between qubit $A$ and the composite
qubits $BC$ quantified by the Wootters' concurrence \cite{wootters1998}. This is to note that tangle is unchanged by permutations of indices that enter in the considered bipartition. For GHZ states, tangle is maximal, i.e., $\tau(|GHZ\rangle)=1$ and it vanishes for all separable, biseparable and W-class of states. Also, tangle does not increase on average under LOCC.\\
The three-tangle has also been proposed for mixed states via convex-roof extension of pure states. For a mixed three-qubit state  $\rho_{ABC} = \sum_i p_i |\psi_i\rangle \langle \psi_i |$, tangle is given as
\begin{eqnarray}
	\tau(\rho_{ABC}) =  min_{\{p_i, |\psi_i\rangle\}}\sum_{i}{p_i \tau(|\psi_i\rangle})
\end{eqnarray} 
where the minimum is taken over all convex decompositions of $\rho_{ABC}$.

\subsection{Partial Tangle}
To further characterize entanglement in three-qubit systems, a quantity known as partial tangle was introduced in \cite{slee2005}. For a three qubit system described by the state $|\psi_{ABC}\rangle$, partial tangles are defined as follows.
\begin{eqnarray}
	\tau_{ij} = \sqrt{C^2_{i|jk} - C^2_{ik}} \label{partangledef}
\end{eqnarray}
for $i \neq j \neq k$ and $i$, $j$, $k$ in $\{A, B, C\}$.\\
Partial tangles help classify the entanglement structure in a three-qubit state \cite{datta2018}:
\begin{itemize}
\item[(i)] If all partial tangles vanish, the state is fully separable.
\item[(ii)] If at least one partial tangle is zero, the state is biseparable.
\item[(iii)] If all are non-zero, the state exhibits genuine tripartite entanglement.
\end{itemize}
If all the partial tangles are zero, then the three qubit state is separable. If atleast one partial tangle is zero, then the state is biseparable. If each partial tangle is non-zero, then the state is genuine entangled \cite{datta2018}.

\subsection{Concurrence fill}
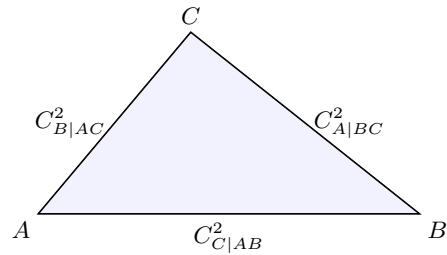
\begin{figure}[h!]
\begin{center}
\begin{tikzpicture}
	\coordinate (A) at (0, 0);
	\coordinate (B) at (5, 0);
	\coordinate (C) at (2, 2.4);
	
	\draw[thick] (A) -- node[midway, below] {$C^2_{C|AB}$} (B)
	-- node[midway, right]  {$C^2_{A|BC}$} (C)
	-- node[midway, left]   {$C^2_{B|AC}$} (A)
	-- cycle;
	
	\node[below left] at (A) {$A$};
	\node[below right] at (B) {$B$};
	\node[above] at (C) {$C$};
	
	 \filldraw[fill=blue!5, draw=black] (A) -- (B) -- (C) -- cycle;
\end{tikzpicture}
\end{center}
	\caption{The above figure shows the concurrence triangle for the three-qubits \textit{A}, \textit{B}, and \textit{C}. The edges of the triangle represents the squared concurrences between bipartitions.} \label{conctri}
\end{figure}

Xie and Eberly introduced a triangle measure of entanglement for three-qubit systems, named as, concurrence fill \cite{eberly2021}. This measure is based on the entanglement polygon inequality which states that one entanglement cannot increase the sum of the other two \cite{qian2018,zhu2015}, i.e.,
\begin{equation}
C^2_{i|jk} \leq C^2_{j|ki} +C^2_{k|ij} \label{ploygon_ineq}
\end{equation}
for any distinct $i$, $j$, $k$ in $\{A, B, C\}$.
A concurrence triangle is the geometrical interpretation of entanglement polygon inequality \cite{eberly2021} (Refer to Fig-\ref{conctri}).

Concurrence fill, denoted by $C_F$, is a genuine tripartite entanglement measure that represents the area of the concurrence triangle.  For a three qubit state, $|\psi_{ABC}\rangle$, concurrence fill is given as
\begin{multline}
	C_F(|\psi_{ABC}\rangle) = \left[\frac{16}{3} Q (Q - C^2_{A|BC}) (Q - C^2_{B|AC}) \right. \\
	\left. \times (Q - C^2_{C|AB}) \right]^{1/4}, \\
	\text{where} \quad Q = \frac{1}{2} \left(C^2_{A|BC} + C^2_{B|AC} + C^2_{C|AB} \right).
	\label{cf}
\end{multline}
$C_F$ is zero for all separable and biseparable states and is non-zero for genuine entangled states (GHZ and W states). It is given as
\begin{eqnarray}
	C_F(|GHZ\rangle)=1 \; \text{and} \; C_F(|W\rangle)=\frac{8}{9}.
\end{eqnarray}
Concurrence fill can be generalized for mixed states via convex roof extension.

\section{Classification of GHZ and W-class of states}
In this section, we delve into the geometric entanglement measure known as concurrence fill for three-qubit pure states. We begin by expressing concurrence fill in terms of the tangle and partial tangle. Subsequently, we introduce a method to classify three-qubit states into the GHZ and W classes using concurrence fill. 

\noindent\textbf{Theorem 1.}
The concurrence fill for a three-qubit state $|\psi_{ABC}\rangle$ can be expressed as follows.
\begin{eqnarray}
	C_F(|\psi_{ABC}\rangle) = \left(\frac{16}{3} \sum_{i\neq j}^{} (\tau_{ij}^2 - \frac{3\tau_{\psi}}{2}) \prod_{i\neq j}^{} (\tau_{ij}^2 - \frac{\tau_{\psi}}{2})\right)^{1/4} \label{area_tau}
\end{eqnarray}
for $i \in \{A,B\}; j \in \{B,C\}$.
\begin{proof}
The partial tangles for the state $|\psi_{ABC}\rangle$ are given as
\begin{eqnarray}
	\tau^2_{ij} = C^2_{ij} + \tau_{\psi}\;\;\; \forall \;\; i,j=A,B,C. \label{tau2ij}
\end{eqnarray}
Using (\ref{tangledef}) in the above equation, we have
\begin{eqnarray}
	C^2_{i|jk} = \tau_{ij}^2 + C^2_{ik} \label{c2ijk}
\end{eqnarray}
for distinct $i$, $j$, $k$ in $\{A,B,C\}$.
Let us consider the semi-perimeter of the concurrence triangle defined in (\ref{cf}).
\begin{eqnarray}
	Q &=& \frac{1}{2} \left(C^2_{A|BC} + C^2_{B|AC} + C^2_{C|AB} \right) \\
	&=& \frac{3}{2} \tau_{\psi} + \left( C^2_{AB}+C^2_{BC}+C^2_{AC} \right) \label{eq18} \\
	&=& \left( \tau^2_{AB}+ \tau^2_{BC}+ \tau^2_{AC} \right) - \frac{3}{2} \tau_{\psi} \label{qtau}
\end{eqnarray}
where (\ref{eq18}) and (\ref{qtau}) follows from (\ref{tau2ij}) and (\ref{c2ijk}).\\
Using (\ref{c2ijk}) and (\ref{qtau}), we have 
\begin{eqnarray}
	Q - C^2_{i|jk} = \tau^2_{jk} - \frac{\tau_{\psi}}{2} \label{q-cijk}
\end{eqnarray}
for distinct $i$, $j$, $k$ in $\{A,B,C\}$.
Using (\ref{qtau}) and (\ref{q-cijk}), the concurrence fill for any three-qubit pure state defined in (\ref{cf}) can be reformulated as
\begin{multline}
		C_F(|\psi_{ABC}\rangle) = \left[ \frac{16}{3}\left(( \tau^2_{AB}+ \tau^2_{BC}+ \tau^2_{AC})  - \frac{3}{2} \tau_{\psi}\right) \right. \\
		\left. \times 
		(\tau^2_{AB} - \frac{\tau_{\psi}}{2})	(\tau^2_{BC} - \frac{\tau_{\psi}}{2})
			(\tau^2_{AC} - \frac{\tau_{\psi}}{2})	
		\right]^{1/4} \label{thm1res}
\end{multline}
which is equivalent to (\ref{area_tau}).
\end{proof}
Now we are in a position to state a few results based on classification of GHZ and W class of states.\\
\noindent\textbf{Theorem 2.} The following inequality holds for W-class of states.
\begin{eqnarray}
	{[C_F(|\psi_{ABC}\rangle)]^4}\geq k\;\tau^2_{BC} \label{kineq}
\end{eqnarray}
where $k = 2(\tau^2_{AB} + C^2_{AB})^{4/3}  (\tau^2_{AC} + C^2_{AC})^{4/3} (\tau^2_{BC} + C^2_{BC})^{1/3}$.
\begin{proof} Let $|\psi_{ABC}\rangle$ be a three-qubit state. Consider the concurrence fill for the state $|\psi_{ABC}\rangle$ defined using tangles and partial tangles in $Theorem- 1$.
Let us first reformulate the following terms in (\ref{thm1res}). Using (\ref{tau2ij}), we obtain
\begin{eqnarray}
	\tau^2_{ij} - \frac{\tau_{\psi}}{2} = C^2_{ij} + \frac{\tau_{\psi}}{2} = \frac{\tau^2_{ij}+ C^2_{ij}}{2} \label{thm2eq1} 
\end{eqnarray}
and applying (\ref{thm2eq1}), we get
\begin{eqnarray}
	( \tau^2_{AB}+ \tau^2_{BC}+ \tau^2_{AC})  - \frac{3}{2} \tau_{\psi} = \sum_{i\neq j}\frac{\tau^2_{ij}+C^2_{ij}}{2} 
	 \label{thm2eq2}
\end{eqnarray}
for $i \in \{A,B\}; j \in \{B,C\}$.\\
Using (\ref{thm2eq1}) and (\ref{thm2eq2}) in (\ref{thm1res}), we have,
	\begin{eqnarray}
	\begin{split}
			C_F(|\psi_{ABC}\rangle) = \bigg(\frac{1}{3} (\tau_{AB}^2 + C^2_{AB}) (\tau_{AC}^2 + C^2_{AC}) \\(2\tau_{BC}^2 - \tau_{\psi})  \sum_{i\neq j}(\tau^2_{ij}+C^2_{ij})\label{area_tau1} \bigg)^{1/4}. 
	\end{split}
	\end{eqnarray}

	Using AM-GM inequality on the three terms in $\sum_{i\neq j}(\tau_{ij}^2 + C^2_{ij})$ in the above equation, we obtain
	\begin{eqnarray}
		\begin{split}
			C_F(|\psi_{ABC}\rangle) \geq ( (\tau_{AB}^2 + C^2_{AB}) (\tau_{AC}^2 + C^2_{AC}) \\(2\tau_{BC}^2 - \tau_{\psi}) \prod_{i\neq j}^{}(\tau^2_{ij} + C_{ij}^2 )^{1/3} )^{1/4}.
		\end{split}
		\label{area_tau2}
	\end{eqnarray}
After re-arranging the terms, the inequality in (\ref{area_tau2}) can be re-expressed as follows.
	\begin{eqnarray}
		\begin{split}
			C_F(|\psi_{ABC}\rangle) \geq (\tau_{AB}^2 + C^2_{AB})^{1/3} (\tau_{AC}^2 + C^2_{AC})^{1/3} \\ (\tau^2_{BC} + C_{BC}^2 )^{1/12} (2\tau_{BC}^2 - \tau_{\psi})^{1/4}
		\end{split}
		\label{ineq1}
	\end{eqnarray}
	Let $|\psi_{ABC}\rangle$ belongs to the W class of state, i.e., $\tau_{\psi} = 0$, and hence the inequality (\ref{ineq1}) is reduced to (\ref{kineq}) which proves the theorem.
\end{proof}
\textbf{Corollary.} Let $|\psi_{ABC}\rangle$ be a three-qubit genuinely entangled state. If the inequality in (\ref{kineq}) is violated, i.e.,
\begin{eqnarray}
	{[C_F(|\psi_{ABC}\rangle)]^4} < k\;\tau^2_{BC} \label{kineq2}
\end{eqnarray}
then,  $|\psi_{ABC}\rangle$ belongs to GHZ class of states.
\begin{example}
	Consider the three-qubit state $|\psi_{cdfe}\rangle = c|001\rangle + d|010\rangle + f|100\rangle + e|011\rangle$, where $|c|^2 + |d|^2 + |f|^2 + |e|^2 = 1$
	\begin{itemize}
		\item[{\textbf{Case 1}.}] For $c=0$, $e=\frac{1}{2}$, $d, f\neq 0$ we have\\
		$|\psi_1\rangle = d|010\rangle + f|100\rangle + \frac{1}{2}|011\rangle$.
		Note that $\tau(|\psi_1\rangle) = e^2f^2 \neq 0$ which implies that $|\psi_1\rangle$ is GHZ state. Now using the criteria given in $Theorem-2$, we see that the inequality (\ref{kineq2}) is not violated (refer to Fig-\ref{figcase1}) which implies that $|\psi_1\rangle$ belongs to GHZ class.
		\begin{figure}[h!]
			\includegraphics[width=0.45\textwidth]{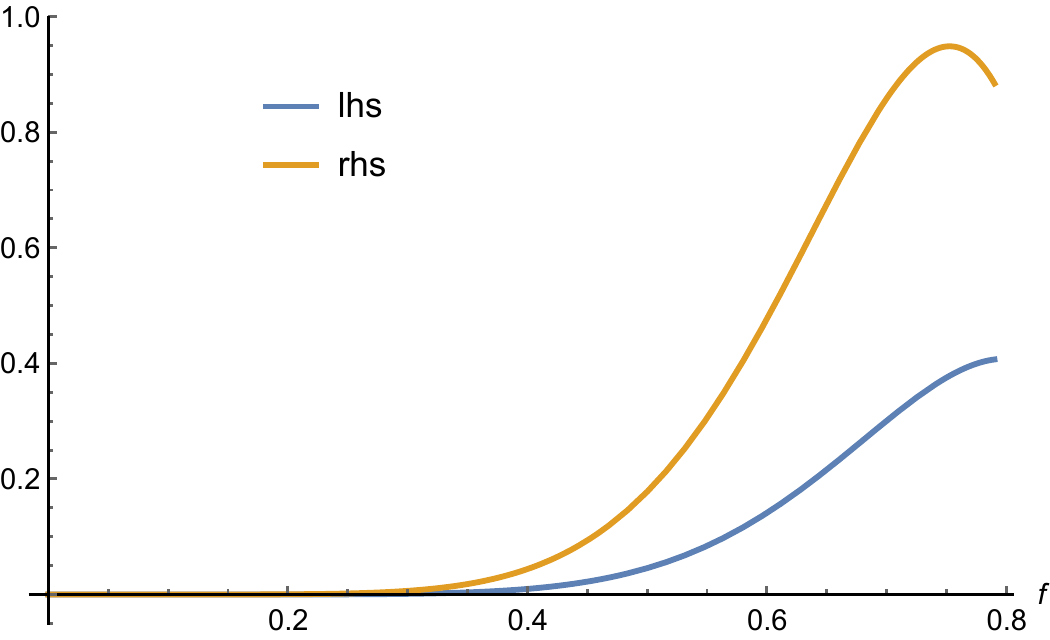}
			\caption{The blue and yellow lines represent the lhs and rhs of the inequality (\ref{kineq2}) for the states $|\psi_1\rangle$. The $x$-axis represents the state parameter $f$.} \label{figcase1}
		\end{figure}
		\item[{\textbf{Case 2}.}] For $e=0$, $c=\frac{1}{\sqrt{5}}$, $d, f\neq 0$, $|\psi_2\rangle = c|001\rangle + d|010\rangle + f|100\rangle$. In this case, we have
		\begin{equation}
			[C_F(|\psi_2\rangle)]^4 \geq k\;\tau^2_{BC}. \label{kineqcase2}
		\end{equation}
		Thus, using (\ref{kineq}), $|\psi_2\rangle$ belongs to W class (Refer to Fig-\ref{figcase2}). Also, since the three-tangle $\tau(|\psi_2\rangle) = e^2f^2 = 0$, and three partial tangles are non-zero, the state $|\psi_2\rangle$ belongs to W class of states.
		\begin{figure}[h!]
			\includegraphics[width=0.45\textwidth]{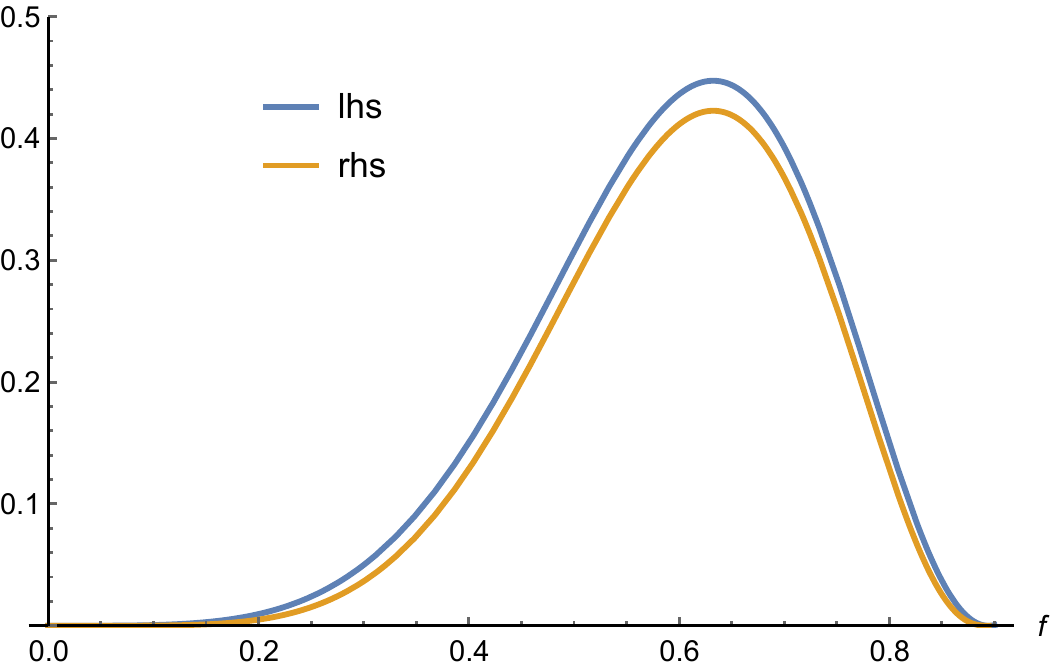}
			\caption{The blue and yellow lines represent the lhs and rhs of the inequality (\ref{kineqcase2}) for the states $|\psi_2\rangle$. The $x$-axis represents the state parameter $f \in [0, \frac{2}{\sqrt{5}}]$.}    \label{figcase2}
		\end{figure}
	\end{itemize}
\end{example}

\section{Concurrence Fill for a Rank-two Mixture of states}
In this section, we study the rank-2 mixture of generalized GHZ and W states. Firstly we derive the minimum value of the concurrence fill for any pure state superposition of generalized GHZ and W states. Next, we formulate the concurrence fill for the mixed states with zero tangle using optimal decomposition method \cite{oster2006,oster2008}.\\
Let us consider the following family of mixed three-qubit states. 
\begin{eqnarray}
	\rho(p) = p |gGHZ\rangle \langle gGHZ| +(1-p)|gW\rangle \langle gW| \label{rhop}
\end{eqnarray}
where $0 \leq p\leq 1$; $|gGHZ\rangle$ and $|gW\rangle$ are the generalized GHZ and W states, respectively, given as follows.
\begin{eqnarray}
	|gGHZ\rangle &=& a|000\rangle +b|111\rangle\\
	|gW\rangle &=& c|001\rangle + d|010\rangle + f|100\rangle 
\end{eqnarray}
where $|a|^2 +|b|^2 = 1$ and  $|c|^2+|d|^2+|f|^2 = 1$.

\subsection{Concurrence fill for eigenstates of $\rho(p)$}
In order to find the concurrence fill for mixed states, let us first study the pure states that are superpositions of the eigenstates of $\rho$. All the elements of the optimal decompositions of $\rho(p)$ are linear combinations of $|gGHZ\rangle$ and $|gW\rangle$. The eigenstates of $\rho(p)$ are given as
\begin{eqnarray}
	|\psi_{p,\phi}\rangle = \sqrt{p} |gGHZ\rangle - \sqrt{1-p}\;e^{i\phi}|gW\rangle. \label{psi}
\end{eqnarray}
In \cite{oster2006}, authors derived the three-tangle for the pure state $|\psi_{p,\phi}\rangle$ which is given as follows:
\begin{eqnarray}
	\tau(|\psi_{p,\phi}\rangle)=4|p^2a^2b^2-4\sqrt{p(1-p)^3}e^{3i\phi}bcdf| \label{taupfi}
\end{eqnarray}
Using the deduction obatined in Appendix, we derive the  concurrence fill of the state $|\psi_{p,\phi}\rangle$ and it is given as follows:
\begin{eqnarray}
	C_F(|\psi_{p,\phi}\rangle) = \left(\frac{256}{3}|xyz(x+y+z)|\right)^{1/4} \label{cfpfi}
\end{eqnarray}
where
\begin{eqnarray}
 	x&=&p^2a^2b^2 + 2p(1-p)e^{2i\phi}f^2b^2 + 2e^{4i\phi} (1-p)^2c^2d^2;\nonumber\\
		y&=&p^2a^2b^2 + 2p(1-p)e^{2i\phi}b^2d^2 + 2e^{4i\phi} (1-p)^2e^2f^2;\nonumber\\
		z&=&p^2a^2b^2 + 2p(1-p)e^{2i\phi}b^2c^2 + 2e^{4i\phi} (1-p)^2d^2f^2.\nonumber\\ \label{xyz}
\end{eqnarray}
We note that 
\begin{eqnarray}
	C_F(|gGHZ\rangle) &=& 4|a^2b^2| \label{cfgghz},\\
	C_F(|gW\rangle) &=& 8|cdf|\left(\frac{|c^2d^2 +d^2f^2+c^2f^2|}{3}\right)^{1/4}. \label{cfgw}
\end{eqnarray}

\begin{figure}[h!]
	\includegraphics[width=0.45\textwidth]{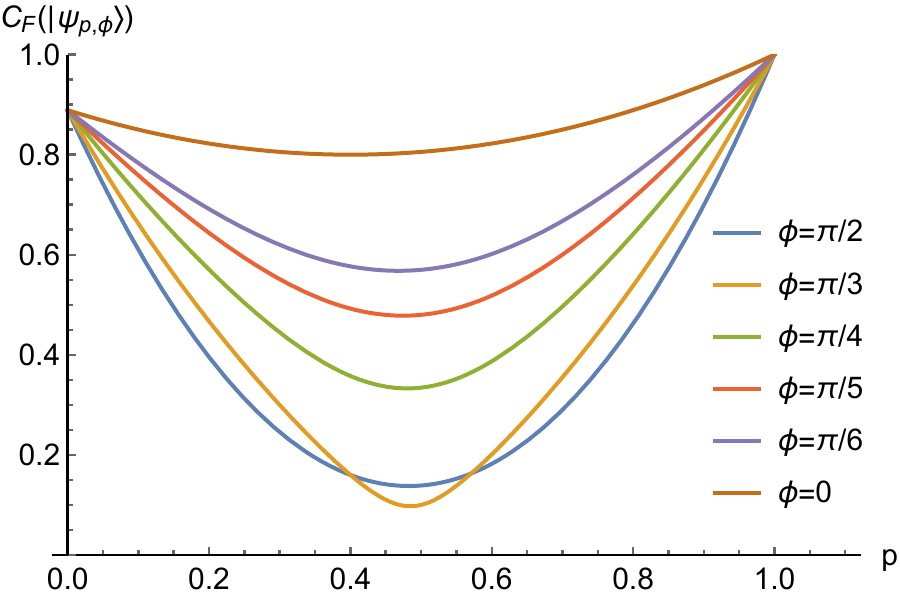}
	\caption{The solid curves represent the concurrence fill of the state $|\psi_{p,\phi}\rangle$ given in (\ref{psi}) for $a=b=\frac{1}{\sqrt{2}}$; $c=d=f=\frac{1}{\sqrt{3}}$ and various values of $\phi$ with $0 \leq p \leq 1$.}
	\label{figcf}
\end{figure}

The concurrence fill $	C_F(|\psi_{p,\phi}\rangle)$ is plotted in Fig-\ref{figcf} for $0\leq p\leq 1$ and different values of $\phi$.
\\

Interestingly, while the expression for concurrence fill obtained in (\ref{cfpfi}) is simpler to use, it depends on the full set of state parameters. To address this, we derive a lower bound for the concurrence fill specifically for the eigenstates of the rank-two mixture $\rho(p)$. This lower bound is particularly useful, as it depends on only a single state parameter, thereby reducing the number of measurements required to estimate the entanglement in the eigenstates of the mixed state.\\
Now let us establish the lower bound of the concurrence fill for the pure eigenstates of the form (\ref{psi}), which is appended in the following theorm.\\
\textbf{Theorem 3.}  For the three-qubit pure state $|\psi_{p,\phi}\rangle$ given in (\ref{psi}), the following inequality holds. 
\begin{eqnarray}
		C_F(|\psi_{p,\phi}\rangle) \geq C_F^{(min)}(|\psi_{p^*,\phi}\rangle) \label{res1} 
\end{eqnarray}
where 
\begin{eqnarray}
	p^* = \begin{cases}
		\frac{1}{1+3b^2}  &\text{for} \;\; n=0,2,4,... \\
		\frac{2+3b^2}{2+15b^2-9b^4}  &\text{for} \;\; n=1,3,5,... \label{pmin}
	\end{cases}
\end{eqnarray} and $C_F^{(min)}(|\psi_{p^*,\phi}\rangle)$
represents the lower bound of the concurrence fill  for the state $|\psi_{p^*,\phi}\rangle$.
\begin{proof} 
Our aim is to find the value of $p$ for which the concurrence fill for the state $|\psi_{p,\phi}\rangle$ attains minima. From (\ref{cfpfi}), we have 
\begin{eqnarray}
	\frac{dC_F(|\psi_{p,\phi}\rangle)}{dp} =\frac{x'}{x} +\frac{y'}{y} +\frac{z'}{z} +\frac{(x'+y'+z')}{x+y+z} \label{dcf}
\end{eqnarray}
where $x'=\frac{dx}{dp},\; y'=\frac{dy}{dp},\; z'=\frac{dz}{dp}$ and $x,y,z$ are defined in (\ref{xyz}).\\
After some calculations, we find that $	\frac{dC_F(|\psi_{p,\phi}\rangle)}{dp} =0$, when 
\begin{eqnarray}
	p=p_1=p_2=p_3 \label{p1p2p3}
\end{eqnarray} 
where
\begin{eqnarray}
	p_1 &=& \frac{2e^{4i\phi}c^2d^2  - e^{2i\phi}f^2b^2}{a^2b^2+2(e^{4i\phi}c^2d^2  - e^{2i\phi}f^2b^2)}\label{p1}\\	p_2&=&\frac{2e^{4i\phi}c^2f^2 -e^{2i\phi}b^2d^2}{a^2b^2+2(e^{4i\phi}c^2f^2-e^{2i\phi}b^2d^2)}\label{p2}\\	
		p_3&=&\frac{2e^{4i\phi}d^2f^2 -e^{2i\phi}b^2c^2}{a^2b^2+2(e^{4i\phi}d^2f^2-e^{2i\phi}b^2c^2)} \label{p3}
\end{eqnarray}
Since $p \in  [0,1]$ is real, we assume the relative phase $\phi=  \frac{n\pi}{2}$ where $n = 0, 1, 2,....$. Here, the $\pi/2$-periodic behavior stems from the relative phase imparted by the local operation $diag\{\exp(i\pi/2), 1\}$ applied individually to every qubit. \\
Now, we find that the condition (\ref{p1p2p3}) holds when $|a|^2 +|b|^2 = 1$ and $c=d=f=\frac{1}{\sqrt{3}}$. The value of $p^*$ satisfying (\ref{p1p2p3}) is obtained as (\ref{pmin}).
Since,  $	\frac{d^2C_F(|\psi_{p,\phi}\rangle)}{dp} \geq 0$ for $|a|^2 +|b|^2 = 1$ and $c=d=f=\frac{1}{\sqrt{3}}$, we conclude that $p^*$ obtained in (\ref{pmin}) is the point of minima with $0.25 < p^* <1$.\\
Hence, we obtain a lower bound of the concurrence fill for the pure states of the form (\ref{psi}) given as
\begin{eqnarray}
	C_F(|\psi_{p,\phi}\rangle) \geq C_F^{(min)}(|\psi_{p^*,\phi}\rangle)  \label{cf_pure}
\end{eqnarray}
where
\begin{eqnarray}
	C_F^{(min)}(|\psi_{p^*,\phi}\rangle) =\begin{cases}
		 \frac{4b^2}{1+ 3b^2} &\text{for} \; n=0,2,4,... \\
		 \frac{4(1-5p^*+4(p^*)^2)}{9} &\text{for}\; n=1,3,5,...
	\end{cases}  \label{cfmin} \; 
	\label{cfminpure}
\end{eqnarray}
with $0 < b < 1$.
\end{proof}
{\example
Consider the symmetric GHZ state and symmetric W state in (\ref{psi}) with $a=b=\frac{1}{\sqrt{2}}$ and $c=d=f=\frac{1}{\sqrt{3}}$. Using (\ref{pmin}), we obtain $p^* = \frac{2}{5}$. Using (\ref{cfminpure}), the concurrence fill is given as $C_F^{(min)}(|\psi_{\frac{2}{5},\phi}\rangle) = \frac{4}{5}$. Hence, the lower bound obtained in (\ref{cf_pure})  conincides with the actual expression of concurrence fill for symmetric mixture of GHZ and W class of states given in (\ref{cfpfi}).

\subsection{Concurrence fill for mixed states}
The convex roof construction is a widely used method for quantifying the entanglement of mixed states.
Let us consider the  mixed state
\begin{eqnarray}
	\rho = \sum_i p_i \Omega_i, \; \Omega_i = \frac{|\phi_i\rangle \langle \phi_i|}{\langle \phi_i |\phi_i \rangle}. \label{omega}
\end{eqnarray}
The concurrence fill of the mixed state $\rho$  can be generalized as convex roof extension of the concurrence fill of pure states,
\begin{eqnarray}
	C_F(\rho) = min_{\{p_i,\Omega_i\}} \sum_i p_i C_F(\Omega_i).
\end{eqnarray}
 The ensemble that minimizes the average value of an entanglement measure under this construction is referred to as the optimal ensemble. While the concurrence of two-qubit mixed states admits an analytical solution, in general, determining the convex roof is highly nontrivial. Nonetheless, several effective approaches have been developed to estimate it, including techniques that provide upper and lower bounds as useful approximations \cite{oster2008, guhne2015}.\\
In \cite{njp2008}, authors provide explicit expression of three-tangle for the generalized mixture of GHZ and W class of states $\rho(p)$ given in (\ref{rhop}). It is shown that upto a certain value of $p \in [0,1]$, the mixed three-tangle vanishes. Hence, there are large number of states in this mixed state family which have vanishing tangle and hence cannot be quantified using three-tangle. Our aim is to quantify the amount of entanglement for the mixed states $\rho(p)$ with zero tangle. We achieve this by estimating the concurrence fill for these states. \\
 In \cite{njp2008}, authors identified the zero polytope containing all the mixed states $\rho(p)$ with zero tangle.
 The optimal decomposition of these states consists of pure states of the form (\ref{psi}) for which three-tangle is zero. It is shown that the $\tau_{\psi} = 0$, for the generalized W state $|gW\rangle$, and three states $|\psi_{p_0,\frac{2k\pi}{3}}\rangle$ for $n=0,1,2$, where
\begin{eqnarray}
	p_0 = \frac{(16c^2d^2f^2)^{1/3}}{(a^4b^2)^{1/3}+(16c^2d^2f^2)^{1/3}} \label{p0}.
\end{eqnarray}
We have
\begin{eqnarray}
	\tau(\rho(p)) = 0 \;\; \text{for} \;\; p \in [0, p_0].
\end{eqnarray}
The optimal decomposition of the zero tangle mixed states $\rho(p)$ is given as follows \cite{njp2008}.
\begin{eqnarray}
\rho(p)=
\frac{p}{p_0} \hat{\rho}(p_0) + \frac{p_o - p}{p_0} \Omega_{gW}; &\; 0 \leq p \leq p_0 \label{opdec}
\end{eqnarray}
where 
\begin{eqnarray}
	\hat{\rho}(p) = \frac{1}{3} \sum_{k=0}^{2} |\psi_{p,\frac{2k\pi}{3}} \rangle \langle \psi_{p,\frac{2k\pi}{3}}| \label{rhopcap}
\end{eqnarray}
with $|\psi_{p,\phi}\rangle$ is defined in  (\ref{psi}) and $\Omega_i$ is defined in (\ref{omega}). \\
Using (\ref{rhopcap}), the concurrence fill of the state $\hat{\rho}(p)$ is given as
\begin{eqnarray}
	C_F(\hat{\rho}(p)) \leq \frac{1}{3}\left[C_F(|\psi_{p,0}\rangle)+C_F(|\psi_{p,\frac{2\pi}{3}}\rangle)+C_F(|\psi_{p,\frac{4\pi}{3}}\rangle)\right]. \label{lbcap}
\end{eqnarray}
Hence, for any mixed state $\rho(p)$ with $\tau(\rho(p))=0$ where $0\leq p \leq p_0$,  the upper bound of the concurrence fill is given as follows.
\begin{eqnarray}
C_F(\rho(p)) \leq C^{ub} _F(\rho(p)) 
\end{eqnarray}
where
\begin{eqnarray}
	 C^{ub} _F(\rho(p)) = \frac{p}{p_0} C_F(\hat{\rho}(p)) + \frac{p_0 -p}{p_0} C_F(|gW\rangle) \label{Cub}
\end{eqnarray}
with $ C_F(\hat{\rho}(p))$ is derived in (\ref{lbcap}) and $C_F(|gW\rangle)$ has been derived in (\ref{cfgw}).

\begin{example}
	Consider the symmetric GHZ and the symmetric W state, i.e., we have $a=b=\frac{1}{\sqrt{2}}$ and $c=d=f=\frac{1}{\sqrt{3}}$ in (\ref{rhop}).
	Using (\ref{p0}), we obtain $p_0 = \frac{4 \sqrt[3]{2}}{3 + 4 \sqrt[3]{2} }$. Using (\ref{Cub}), the upper bound of the concurrence fill for generalized mixture of states represented by $\rho(p)$ for $p \in [0, p_0]$ is given in Fig-\ref{fig4}.
		\begin{figure}[h!]
		\includegraphics[width=0.45\textwidth]{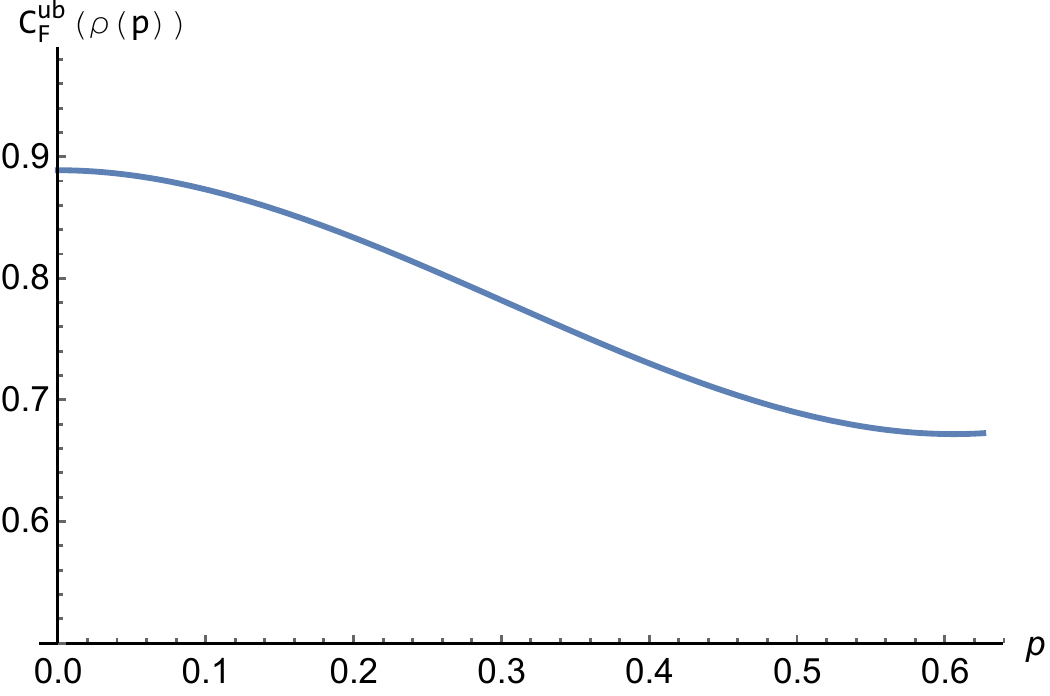}
		\caption{The blue curve represents the upper bound of the concurrence fill given in (\ref{Cub}) for the zero-tangle states in the generalized mixture $\rho(p)$ with $0 \leq p \leq p_0$.}
		\label{fig4}
	\end{figure}
\end{example}
\section{Conclusion}
In this work, we addressed the challenge of detecting and classifying three qubit genuine entangled states, which remains elusive under many conventional measures despite experimental realizations.  To achieve the task, we study the concurrence fill for pure as well as mixed three-qubit states. We began by reformulating the concurrence fill in terms of three tangle and partial tangle. This characterization is important because tangle is LOCC monotonic and partial tangle can be measured using expectation value of measurement operators which makes them operationally meaningful. We further employ our formulation to classify GHZ and W class of states. Since convex roof extension is a computationally challenging task, we applied our framework to analyze a rank-2 mixture of generalized GHZ and W-class states, which serves as a representative for nontrivial mixed quantum states. For the eigenstates of the resulting generalized mixed states, we derived an explicit expression for concurrence fill and established an upper bound for the zero tangle mixed states.
These findings not only enhance the utility of concurrence fill as a diagnostic tool for detecting genuine three qubit entangled states, but also open the door to future studies on geometric characterizations of quantum measures in multipartite systems.

\section{Acknowledgement}
The author acknowledges initial academic guidance provided by Dr. Satyabrata Adhikari during the early stages of this work.
\section{DATA AVAILABILITY STATEMENT}
Data sharing not applicable to this article as no datasets were generated or analysed during the current study.
\section{Appendix}
Consider the pure three-qubit state  $|\psi_{ABC}\rangle$. Let $\rho_{AB}$ and $\rho_{AC}$ denote the density matrices representing the reduced states of the three-qubit state $\rho_{ABC} = |\psi_{p,\phi}\rangle \langle \psi_{p,\phi}|$. Let $\tilde{\rho}_{ij}$ be the spin flipped density matrix as defined in (\ref{rholtilde}) and $\lambda_{n}^{ij} $ be the square root of the $n^{th}$ eigenvalue of the state $\rho_{ij}{\tilde{\rho}}_{ij}$, where $i,j$ are the distinct index representing the subsystems $A,B,C$ and eigenvalues are arranged in decreasing order.
 In \cite{ckw2000}, authors defined the concurrence of the reduced two-qubit state as follows:
\begin{eqnarray}
	C_{ij} = (\lambda_1^{ij}-\lambda_2^{ij}) \; \text{for}\; i \neq j. \label{cij}
\end{eqnarray}
The three-tangle defined in (\ref{tangledef}) can be defined in terms of the eigenvalues of reduced states as follows \cite{ckw2000}.
\begin{eqnarray}
	\tau(|\psi_{p,\phi}\rangle)= 2(\lambda_1^{AB}\lambda_2^{AB} + \lambda_1^{AC}\lambda_2^{AC} ) \label{taulambda}
\end{eqnarray}
Using (\ref{cij}) and (\ref{taulambda}) in (\ref{tangledef}), we obatin
\begin{eqnarray}
	C^2_{A|BC} &=& 	\tau(|\psi_{p,\phi}\rangle) + C^2_{AB} +  C^2_{AC} \nonumber\\
&=& 2(\lambda_1^{AB}\lambda_2^{AB} + \lambda_1^{AC}\lambda_2^{AC} ) \nonumber \\  &&+(\lambda_1^{AB}-\lambda_2^{AB})^2 + (\lambda_1^{AC}-\lambda_2^{AC})^2 \nonumber \\
&=& {\lambda_1^2}^{(AB)} +  {\lambda_2^2}^{(AB)} + {\lambda_1^2}^{(AC)} +  {\lambda_2^2}^{(AC)} \label{c2abc}
\end{eqnarray}
Similarly, we have 
\begin{eqnarray}
	C^2_{B|AC} &=&  {\lambda_1^2}^{(AB)} +  {\lambda_2^2}^{(AB)} + {\lambda_1^2}^{(BC)} +  {\lambda_2^2}^{(BC)} \label{c2bac}
\end{eqnarray}
and 
\begin{eqnarray}
	C^2_{C|AB} &=&  {\lambda_1^2}^{(BC)} +  {\lambda_2^2}^{(BC)} + {\lambda_1^2}^{(AC)} +  {\lambda_2^2}^{(AC)} \label{c2cab}
\end{eqnarray}
Using above equations (\ref{c2abc})-(\ref{c2cab}) in (\ref{cf}), we obtain 
\begin{eqnarray}
	Q = \sum_{k=1,2} ({\lambda_k^2}^{(AB)} +  {\lambda_k^2}^{(BC)} +  {\lambda_k^2}^{(AC)}) \label{qeigen}
\end{eqnarray}
and hence the concurrence fill is obtained as 
\begin{multline}
	C_F(|\psi_{ABC}\rangle) = \bigg( \frac{16}{3} \sum_{k=1,2} ({\lambda_k^2}^{(AB)} +  {\lambda_k^2}^{(BC)} +  {\lambda_k^2}^{(AC)})\\ \times ({\lambda_1^2}^{(AB)} +  {\lambda_2^2}^{(AB)}) ({\lambda_1^2}^{(BC)} +  {\lambda_2^2}^{(BC)}) ({\lambda_1^2}^{(AC)} +  {\lambda_2^2}^{(AC)}) \bigg)^{1/4}. \label{cfeigen}
\end{multline}


\newpage
\begin{thebibliography}{90}
	\bibitem{karlson1998} A. Karlsson, M. Bourennane, Phys. Rev. A\textbf{ 58}, 4394 (1998).
	\bibitem{yeo2006} Y. Yeo, and W. K. Chua, Phys. Rev. Lett. \textbf{96}, 060502 (2006).
	\bibitem{hillery1999} M. Hillery, V. Bužek and A. Berthiaume, Phys. Rev. A \textbf{59}, 1829 (1999).
		\bibitem{wootters1998} W. K. Wootters, Phys. Rev. Lett. \textbf{80}, 2245 (1998).
	\bibitem{wootters1997} S. Hill and W. K. Wootters, Phys. Rev. Lett. \textbf{78}, 5022 (1997).
	\bibitem{vidal2002} G. Vidal, R.F. Werner, Phys. Rev. A \textbf{65}, 032314 (2002).
	\bibitem{carvalho2004}	A.R.R. Carvalho, F. Mintert, A. Buchleitner,
	Phys. Rev. Lett. \textbf{93}, 230501 (2004).
	\bibitem{jungnitch2011}	B. Jungnitsch, T. Moroder, O. Gühne, Phys. Rev.
	Lett. \textbf{106} 190502 (2011)
	\bibitem{vidal2001} G. Vidal, R. Tarrach, Phys. Rev. A \textbf{59}, 141 (1999).
	\bibitem{acin2001} A. Acin, A. Andrianov, E. Jane and R Tarrach, J. Phys. A: Math. Gen. \textbf{34}, 6724 (2001).
	\bibitem{acinprl2001} A. Acin, D. Bruss, M. Lewenstein, and A. Sanpera, Phys. Rev. Lett. \textbf{87}, 4 (2001).
	\bibitem{dur2000} W. Dur, G. Vidal, and J. I. Cirac, Phys. Rev. A \textbf{62}, 062314 (2000).
   \bibitem{ckw2000} V. Coffman, J. Kundu, and W.K. Wootters, Phys. Rev. A \textbf{61}, 052304 (2000).
	\bibitem{qaisar2017} S. Qaisar,  J. U. Rehman, Y. Jeong, H. Shin, Prog Theor Exp Phys \textbf{2017(4)}, 041A01 (2017).
	\bibitem{sen2016} A. Kumar, S. S. Roy, A. K. Pal, R. Prabhu, A. S. De, U. Sen, Phys. Rev. A \textbf{380}, 3588 (2016)
	
	\bibitem{verstrate2006}  T.J. Osborne, F. Verstraete, Phys. Rev. Lett. \textbf{96(22)}, 220503 (2006)
	\bibitem{olive2014} T. R. de Oliveira, M. F. Cornelio, F. F. Fanchini, Phys. Rev. A \textbf{89(3)}, 034303 (2014)
	\bibitem{fan2007} Y. C. Ou, H. Fan,  Phys. Rev. A \textbf{75(6)}, 062308 (2007)
	\bibitem{kim2009} J. S. Kim, A. Das, B. C. Sanders, Phys. Rev. A \textbf{79(1)}, 012329 (2009)
	\bibitem{kim2010} J. S. Kim, Phys. Rev. A \textbf{81(6)}, 062328 (2010)
	\bibitem{luo2016} Y. Luo, T. Tian, L. H. Shao, Y. Li, Phys. Rev. A \textbf{93(6)}, 062340 (2016)
	\bibitem{sanders2010} J. S. Kim, B. C. Sanders,  J. Phys. A: Math. Theor. \textbf{43(44)}, 445305 (2010)

	
	\bibitem{ma2011} Z.-H. Ma, Z.-H. Chen, J.-L. Chen, C. Spengler, A. Gabriel,
	and M. Huber, Phys. Rev. A \textbf{83}, 062325 (2011).
\bibitem{eberly2021} S. Xie and J. H. Eberly, Phy. Rev. Lett. \textbf{127}, 040403 (2021).
\bibitem{li2022} Y.-W. Li, L.-J. Li, X.-K. Song, et al., Eur. Phys. J. C 82, 799 (2022).
\bibitem{mishra} A. Mishra, S. Mahanty, A. K. Roy, P. K. Panigrahi, Physics Open \textbf{20}, 100230 (2024).
\bibitem{Ge2023} X. Ge, L. Liu,2, and S. Cheng, Phys. Rev. A \textbf{107}, 032405 (2023)
\bibitem{njp2008} C. Eltschka, A. Osterloh, J. Siewert and A. Uhlmann, New J. Phys. \textbf{10}, 043014 (2008).
\bibitem{datta2018} C. Datta, S. Adhikari, A. Das, P. Agrawal, Eur. Phy. J. D \textbf{72}, 157 (2018).
\bibitem{vedral}  V. Vedral, M.B. Plenio, M.A. Rippin, et al., Phys. Rev. Lett. \textbf{78}  2275 (1997).
\bibitem{hororev2009} R. Horodecki, P. Horodecki, M. Horodecki, and K. Horodecki, Rev. Mod. Phys. \textbf{81}, 865 (2009)
\bibitem{aggarwal2021} S. Aggarwal, S. Adhikari, Quantum Inf Process \textbf{20}, 83 (2021)
\bibitem{ligao2024} H. Li, T. Gao, F. Yan,  Phys. Rev. A \textbf{109}, 012213 (2024)
\bibitem{zhu2020} X. N. Zhu, M. J. Zhao, S.M. Fei, Int. J. Theor. Phys. \textbf{59}, 1688 (2020)
\bibitem{cornelio2013} M. F. Cornelio, Phys. Rev. A \textbf{87}, 032330 (2013)
\bibitem{slee2005} S. Lee, J. Joo, J. Kim, Phys. Rev. A \textbf{72}, 024302 (2005). 
\bibitem{qian2018} X.-F. Qian, M. A. Alonso, and J. H. Eberly, New J. Phys. \textbf{20}, 063012 (2018).
\bibitem{zhu2015} X.-N. Zhu and S.-M. Fei, Phys. Rev. A \textbf{92}, 062345 (2015).

\bibitem{oster2006} R. Lohmayer, A. Osterloh, J. Siewert, and A. Uhlmann, Phys. Rev. Lett. \textbf{97}, 260502 (2006).
\bibitem{oster2008} A. Osterloh, J. Siewert and A. Uhlmann,  Phys. Rev. A \textbf{77} 032310 (2008)

\bibitem{guhne2015} G. Toth, T. Moroder, O. Guhne, Phys. Rev. Lett \textbf{114}, 160501 (2015)
\end{thebibliography}
\end{document}